\documentclass{JHEP3}
\usepackage{amsmath, amsfonts}
\usepackage{epsfig}
\usepackage{cite}

\providecommand{\openone}{\leavevmode\hbox{\large1\kern-7.3pt\normalsize1}}
\newcommand{\be}{\begin{equation}}
\newcommand{\ee}{\end{equation}}
\newcommand{\ba}{\begin{eqnarray}}
\newcommand{\ea}{\end{eqnarray}}

\newcommand{\nn}{\nonumber \\}
\newcommand{\fr}[2]{{\frac{#1}{#2}\,}}

\renewcommand{\vec}[1]{{\bf #1}}
\renewcommand{\(}{\left(}
\renewcommand{\)}{\right)}

\def\be{\begin{equation}}
\def\ee{\end{equation}}

\def\calj{{\mathcal Q}}

\preprint{arXiv.org/0708.0609 [hep-th]}

\title
    {%
    Spinning Dragging Strings
     }

\author
    {%
    C.~P.~Herzog$^1$\footnote{\tt herzog@phys.washington.edu}~ and A.~Vuorinen$^{12}$\footnote{\tt vuorinen@phys.washington.edu}    \\
   $^1$Department of Physics, University of Washington, Seattle, WA   98195--1560\\
   $^2$Institut f\"ur Theoretische Physik, Technische Universit\"at Wien, Wiedner Hauptstr. 8-10, A-1040 Vienna, Austria

    }

\abstract{
We use the AdS/CFT correspondence to
compute the drag force experienced by a heavy quark
moving through a ${\mathcal N}=4$ $SU(N)$ super Yang-Mills plasma at nonzero
temperature and R-charge chemical potential and
at large 't Hooft coupling.
We resolve a discrepancy in the literature between two earlier studies of
such quarks.  In addition, we consider small fluctuations of the spinning strings
dual to these probe quarks and find no evidence of instabilities.  We make
some comments about suitable D7-brane boundary conditions for the dual strings.
}


\begin{document}

\section{Introduction and Summary}

We address an apparent discrepancy between two AdS/CFT models
\cite{Herzog:2006se,Caceres:2006dj}
of heavy quark energy loss in ${\mathcal N}=4$ $SU(N)$ super Yang-Mills (SYM)
theory at nonzero R-charge chemical potential.
Both claim to study the drag force felt by a heavy probe quark of a
${\mathcal N}=2$ hypermultiplet in the limit of strong 't Hooft coupling
as a function of velocity, temperature, and R-charge.
Building on results of Refs.~\cite{Herzog:2006gh, Gubserdrag},
both claim the behavior of the quark
can be described by the simple equation
\be
\frac{dp}{dt} = - \mu p \ ,
\ee
where $p$ is the momentum and $\mu$ the friction coefficient.
However,
Ref.~\cite{Caceres:2006dj} finds the friction coefficient $\mu$
is velocity independent while the corresponding $\mu$ of Ref.~\cite{Herzog:2006se}
is velocity dependent.

The quarks are modeled by strings in supergravity backgrounds,
and at the time these papers appeared, the discrepancy was attributed to the fact that
Ref.~\cite{Caceres:2006dj} used the full ten dimensional supergravity solution while
Ref.~\cite{Herzog:2006se} used a five dimensional truncation.  In this paper, we find
a large family of solutions in the ten dimensional supergravity background.  The
single charge solutions of Ref.~\cite{Caceres:2006dj} belong to this family.
There is also a sense in which
a ten dimensional uplift of the solutions found in Ref.~\cite{Herzog:2006se} belong
to this family.

The maximally supersymmetric
${\mathcal N}=4$ SYM has an $SO(6)$ R-symmetry group, and we are free
to introduce a chemical potential dual to any of the three elements of the
Cartan sub-algebra of $SO(6)$.  While the gravity dual
to ${\mathcal N}=4$ SYM at zero temperature and zero chemical potential
is the $AdS_5 \times S^5$ background, the gravity dual at nonzero
temperature and chemical potential is the near horizon
limit of a spinning
D3-brane background
\cite{Gubserspinning,Behrndt,Cai:1998ji,Chamblin:1999tk,CveticGubser}.
This background consists of a Kerr-type
black hole with $AdS_5 \times S^5$ asymptotics, where the
rotation is in the $S^5$ directions.  The $S^5$ geometrically realizes
the $SO(6)$ R-symmetry.

To add a hypermultiplet, Ref.~\cite{KatzKarch}
demonstrated that one adds a D7-brane to the $AdS_5 \times S^5$
geometry that wraps an $S^3 \subset S^5$.  The D7-brane
also wraps $AdS_5$ down to some minimal radius $r_0$
that roughly speaking plays the role of the quark mass.
At finite temperature and chemical potential, the story is roughly
the same although the details of the embedding are sensitive
to the geometry.  Single quark solutions correspond to strings
that stretch from the D7-brane to the horizon.

We work in the large $N$ limit where it is consistent
to ignore the effect of the string on the D7-brane and also to ignore
the effect of the string and the D7-brane on the black hole geometry.
The Nambu-Goto
action of the string scales as $\sqrt{\lambda}$ where $\lambda = g_{YM}^2 N$ is
the 't Hooft coupling, while the DBI action for a single D7-brane scales as
$N \lambda$.  We work in the limit $N \to \infty$, where $\lambda$ is kept
large and fixed.  Thus, we expect the string to affect the D7-brane only at order
$1/N$.  The supergravity action scales as $N^2$, and thus the effect of the
string and the D7-brane on the geometry is also suppressed by powers of $1/N$.

The addition of the D7-brane breaks the $SO(6)$ R-symmetry
down to an $SO(4)$ subgroup.
Thus, at finite R-charge chemical
potential, the physics of the dragging quarks will be sensitive
to the matching between the unbroken $SO(4)$ and the
chemical potentials that we decide to turn on.
There are two simple cases to consider.

If we turn on a chemical potential for an element
of the Cartan subalgebra inside the unbroken $SO(4)$, the
D7-brane and string will be uncharged with respect to the
unbroken $SO(4)$ and should be affected by the chemical potential
only indirectly through the effect of the potential on the metric.
To visualize a cartoon of the situation, imagine the rotating black hole as
an $S^2$ spinning in the $xy$-plane.  Above the north pole of the rotating
$S^2$, we place the D7-brane, which we imagine as a two dimensional
plane parallel to the $xy$-plane.  The string solution stretches along
the $z$-axis from the D7-brane to the north pole of the black hole.
Ref.~\cite{Caceres:2006dj} looked at strings
that stretch from the black hole horizon to the boundary of the space time which
correspond to infinitely massive quarks.  However, it
is straightforward to cut off their strings at some finite radius by adding a D7-brane,
and indeed
the string just discussed is precisely the regulated polar string of Ref.~\cite{Caceres:2006dj}.
A cartoon of such a string is drawn in Figure \ref{stringcartoon}a.

In contrast, if we turn on a chemical potential for the element of the
Cartan subalgebra in the broken part of $SO(6)$, we expect the
string to be directly affected by the chemical
potential.  Now the D7-brane becomes a line extending in the $z$ direction
in the cartoon at a fixed
value of $x$ and $y$.
The string spirals around the black
hole in the $xy$-plane.  Such a string is precisely the equatorial
string of Ref.~\cite{Caceres:2006dj}.
In this second case, the string will exert a torque on the D7-brane.
As emphasized above,
we work in the large $N$ limit where the D7-brane is far more massive than
the string, and we ignore the backreaction of the string on
the D7-brane.  Cartoon cross sections of this string are shown as Figures
\ref{stringcartoon}b and \ref{stringcartoon}c.

\FIGURE{
\centerline{\raisebox{3cm}{(a)} \psfig{figure=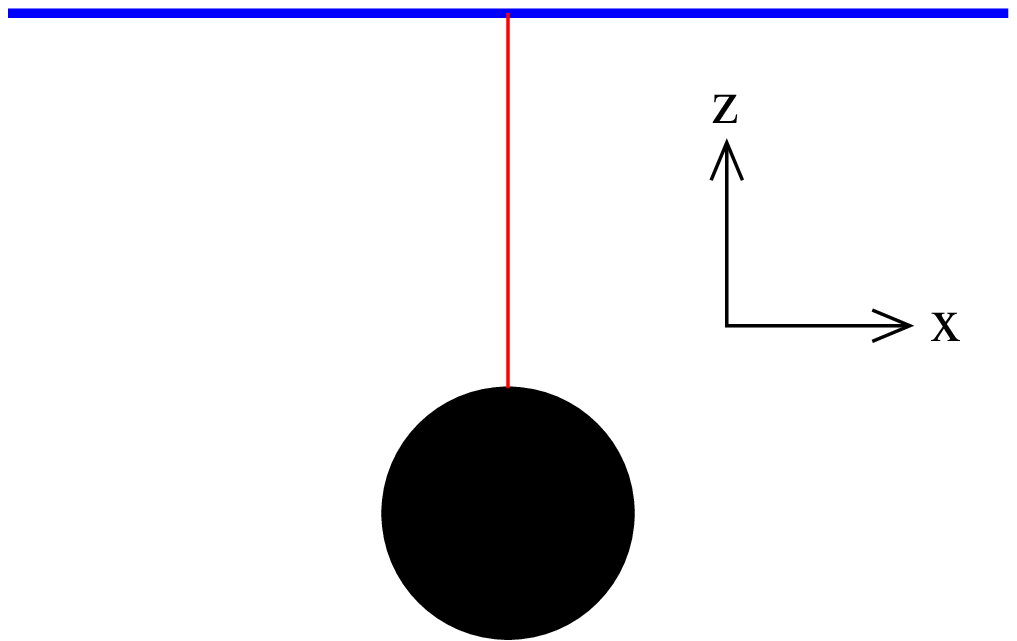, width=2.3in}
\raisebox{3cm}{(b)}
 \psfig{figure=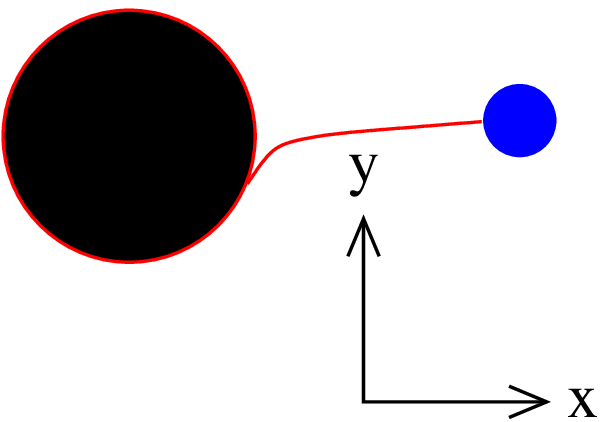, width=1.5in}
\raisebox{3cm}{(c)}
\psfig{figure=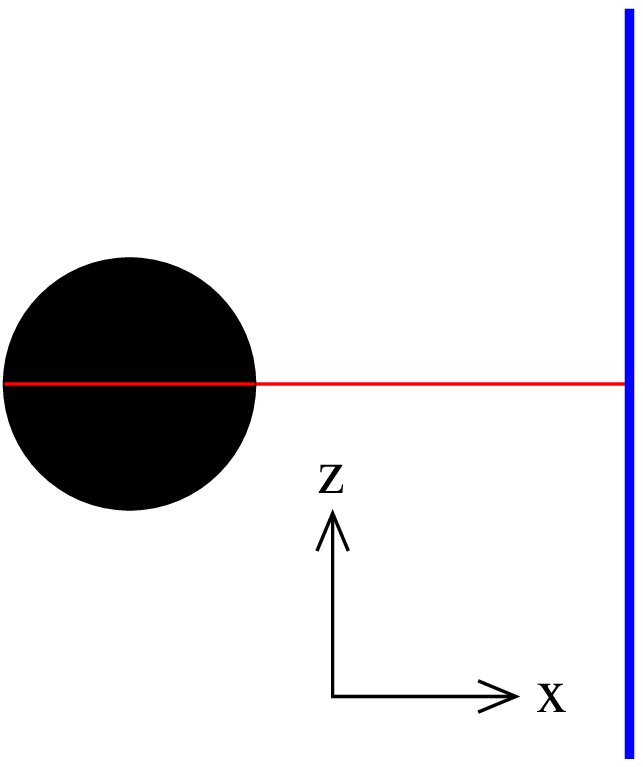, width=1.5in}}
\caption{Cartoons of the spinning strings.  The black disk is the
black hole, the thin red line is the string, and the thick blue line or dot
is the D7-brane: a) is the polar string; b) and c) are views
of the equatorial spinning string.}
\label{stringcartoon}
}

In this paper, we describe a family of solutions, which interpolate between the equatorial
strings of Ref.~\cite{Caceres:2006dj} and the uplifted strings of
Ref.~\cite{Herzog:2006se}.
The strings of Ref.~\cite{Herzog:2006se} have a more complicated field theory
interpretation.
The uplift of the 5d string of
Ref.~\cite{Herzog:2006se} spins in the $S^5$ directions but requires no torque,
while the equatorial
string of Ref.~\cite{Caceres:2006dj} requires a torque to be applied in the
transverse $S^5$ directions but does not spin in these directions,
In Section \ref{sec:D7necks}, we construct
a D7-brane configuration that would allow the no-torque strings of Ref.~\cite{Herzog:2006se} to spin.
A cartoon of the configuration is presented as Figure \ref{stringcartoonfancy}.
This necked-configuration, like the simpler D7-branes considered above, breaks the $SO(6)$
R-symmetry down to $SO(2)\times SO(4)$.  
Unfortunately, the D7-brane appears to be unstable.

As has been implicit in the discussion up to now, it is important to distinguish between the
$SO(6)$ R-symmetry breaking due to nonzero chemical potential from the breaking due to including
D7-branes.  We consider string solutions in this paper where all three R-charge chemical
potentials are nonzero and the R-symmetry is broken to $SO(2)^3 \cong U(1)^3$.  
We also consider string solutions where only one R-charge chemical potential is nonzero 
and the breaking pattern is $SO(4) \times SO(2)$.
From the gravitational
point of view, chemical potential is dual to the rotation of a black hole; 
at zero chemical potential, the R-symmetry is manifest as the symmetry group of an $S^5$ which
at nonzero chemical potential gets squashed by the effects of angular rotation.  
The D7-branes, or boundary conditions, we consider, by contrast, only break $SO(6)$ to $SO(4)\times SO(2)$, a breaking
which may or may not be compatible with the breaking from the chemical potential.  
D7-brane configurations with a more elaborate
symmetry breaking pattern are outside the scope of this paper, but they may well exist,
and they may well have an interesting field theory meaning. 

\FIGURE{
\centerline
{\psfig{figure=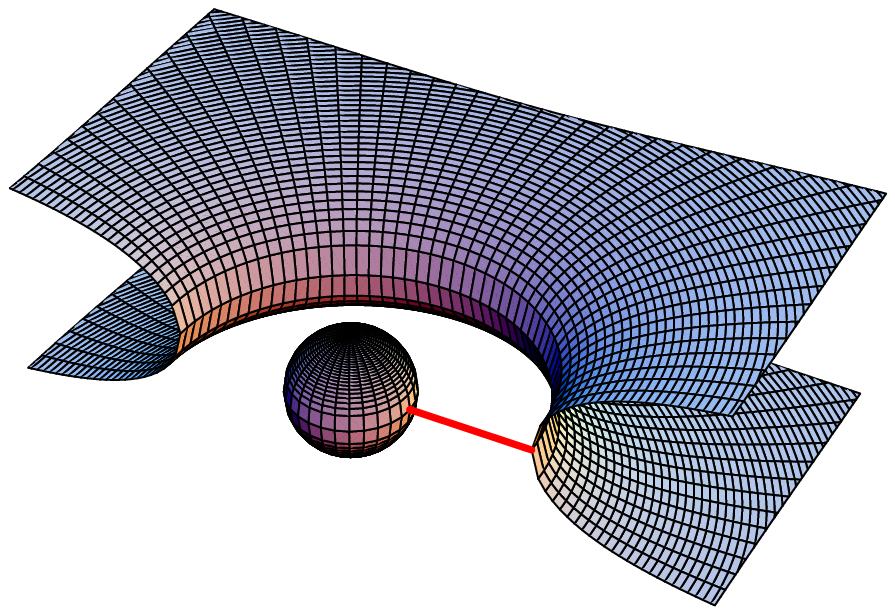, width=3.2in}
}
\caption{Cartoon of the proposed D7-brane
configuration for the torqueless string.  The
string is caricatured by a red line segment.}
\label{stringcartoonfancy}
}

We begin in Section \ref{sec:metric} by reviewing the metric describing the near horizon
limit of spinning D3-branes in ten dimensional type IIB supergravity.
Next, in Section \ref{sec:generalspinning}, we construct a family of spinning string
solutions that interpolates between the equatorial strings of Ref.~\cite{Caceres:2006dj} and the
no-torque strings.
In Section \ref{sec:fived}, we demonstrate how the five dimensional strings of
Ref.~\cite{Herzog:2006se} uplift to the no-torque strings in ten dimensions.
Then, in Section \ref{sec:D7necks}, we discuss what types of D7-branes
our spinning strings can end on.
In Section \ref{sec:stability}, we consider the stability of the various spinning string solutions
against small fluctuations.
A Discussion concludes with some comments about universality and relevance
of these solutions for understanding energy loss of quarks in QCD.

\section{The spinning black D3-brane background}
\label{sec:metric}

Here we reproduce the metric for the near horizon limit of the spinning D3-brane background
(see for example \cite{Behrndt}).
We take the black hole horizon to have translational symmetry in the gauge theory directions,
i.e.~we are working in a gauge theory on Minkowski space ${\mathbb R}^{3,1}$.
The metric is as follows,\footnote{In choosing the sign of the $A^i$ term here, we are following the conventions of Ref.~\cite{Cvetic:1999xp}.}
\be
\label{D3metric}
ds_{10}^2 = \sqrt{\Delta} ds_5^2 + \frac{R^2}{\sqrt{\Delta}} \sum_{i=1}^3 X_i^{-1} \left( d\mu_i^2 + \mu_i^2 (d \psi_i + A^i/R)^2 \right) \ ,
\ee
where
\be
\label{fivered}
ds_5^2 = -(H_1 H_2 H_3)^{-2/3} h \, dt^2 + (H_1 H_2 H_3)^{1/3}(h^{-1} dr^2 + \frac{r^2}{R^2} d \vec x^2) \ ;
\ee
\be
X_i = H_i^{-1} (H_1 H_2 H_3)^{1/3} \; ; \; \; \; A^i = \frac{(1-H_i^{-1}) \sqrt{m}}{\ell_i} dt \ ;
\ee
\be
h = - \frac{m}{r^2} + \frac{r^2}{R^2} H_1 H_2 H_3 \; ; \; \; \;
\Delta = \sum_{i=1}^3 X_i \mu_i^2 \; ; \; \; \;
H_i = 1 + \frac{\ell_i^2}{r^2} \ ;
\ee
\be
\mu_1 = \sin \theta_1 \; ; \; \; \; \mu_2 = \cos \theta_1 \sin \theta_2 \; ; \; \; \;
\mu_3 = \cos \theta_1 \cos \theta_2 \ .
\ee
There is also an RR five-form flux $F_5$ which does not affect the string dynamics, so we will
ignore it.
At large $r$, the five dimensional metric (\ref{fivered}) asymptotically approaches that
of $AdS_5$.
The constant $R$ is the radius of curvature of this asymptotically $AdS_5 \times S^5$
spacetime.

By making the $\ell_i$ nonzero, we introduce an R-charge chemical potential.  From the
gravitational point of view, we have a black hole of the Kerr type with angular momentum
in the $S^5$ directions determined by the $\ell_i$.
The horizon radius is determined by the equation $h(r_h)=0$.
There is also an ergoregion, the boundary of which is determined by the vanishing of $g_{tt}$.
Our strategy in this paper will be to avoid working directly with this metric until the last
possible moment.

To give the reader a better feel for the field theory side of the story,
we review some elements of the AdS/CFT dictionary that
translate these gravitational quantities into their field theory counterparts.
First recall that the 't Hooft coupling is $g_{YM}^2 N = \lambda = R^4 / \alpha'^2$
where $1 / 2 \pi \alpha'$ is the string tension.
The Hawking temperature of the black hole, which is also the temperature of the field theory, is
\[
T = \frac{1}{2 \pi r_h^2 R^2}
\frac{2 r_h^6 + (\ell_1^2+\ell_2^2+\ell_3^2) r_h^4 - \ell_1^2 \ell_2^2 \ell_3^2}
{\prod_{i=1}^3 \sqrt{r_h^2 + \ell_i^2}} \ .
\]
The chemical potentials are directly related to the $A^i$.  In fact, had we subtracted
a constant term such that $A^i$ was zero on the black hole horizon, then we could read
off the chemical potential by evaluating $A^i$ at the boundary $r \to \infty$.  Instead,
we need to subtract off the horizon contribution, yielding
\[
\Phi_i = \left. \frac{A^i}{R} \right|_{r \to \infty} - \left. \frac{A^i}{R} \right|_{r=r_h} =
-\frac{\ell_i \prod_{j=1}^3 \sqrt{r_h^2+ \ell_i^2}}{R^2 r_h (r_h^2 + \ell_i^2)} \ .
\]
Note that the chemical potential $\Phi_i$ vanishes if $\ell_i=0$.

\section{General spinning strings}
\label{sec:generalspinning}

Keeping track of the full spinning D3-brane metric is a complicated business, and
to find interesting spinning string solutions, we need only focus on a six dimensional
slice of the full ten dimensions.  In particular, we are going to ignore the $\theta_a$
directions of (\ref{D3metric}) because it simplifies the analysis.  At the end of the
day, the family of string solutions we find that
interpolates between the equatorial strings of Ref.~\cite{Caceres:2006dj} and the uplifted strings
of Ref.~\cite{Herzog:2006se} has only a trivial $\theta_a$ dependence.
Consider therefore a line element of the form
\be
ds^2 = f \left[ - \alpha \, dt^2 + \beta \, dr^2 + \gamma \, d  x^2 \right] +
\frac{1}{f} \sum_i \epsilon_i
(d\psi_i + \phi_i \, dt)^2 \ .
\label{genericmetric}
\ee
The variables $f$, $\alpha$, $\beta$, $\gamma$, $\epsilon_i$, and $\phi_i$ are
functions of the radial coordinate $r$, and their values can be read off from the results displayed in the previous
Section; accordingly, $f$ and $\epsilon_i$ also depend on auxiliary angular coordinates $\theta_a$.
We will think of this line element as a slice of the spinning
D3-brane metric, where we have set the polar angles $\theta_a$ to constant values.
The equations of motion will typically be satisfied only for special values of the
$\theta_a$.

We will look for stationary spinning string solutions that can stretch from the boundary
of the space-time to the black hole horizon.
We derive the equations of motion for the string from the Nambu-Goto action
\be
S = -\frac{1}{2\pi \alpha'} \int d\sigma \, d \tau \, \sqrt{- \det G}
\ee
where $G_{ab}$ is the induced metric on the string world-sheet.
We take a static gauge where
$\tau = t$, $\sigma = r$, and the string extends in the four directions $x(\sigma, \tau)$,
$\psi_i(\sigma, \tau)$.  We look for stationary solutions and so assume that
$\dot x = v$ and $\dot \psi_i = \omega_i$ are constant.
Defining $X = (t, r, x, \psi_1, \psi_2, \psi_3)$
and $U \cdot V = U^\mu V^\nu g_{\mu\nu}$, where $g_{\mu\nu}$
is the space-time metric, we find that
\be
{\mathcal L}^2 \equiv -\det G = (\dot X \cdot X')^2 - (X')^2 (\dot X)^2 \ ,
\label{calLdef}
\ee
where
\begin{eqnarray}
(\dot X)^2 &=& -\alpha f + f \gamma v^2 + \frac{1}{f} \sum_i \epsilon_i (\phi_i + \omega_i)^2 \ , \\
(X')^2 &=& \beta f + f \gamma (x')^2 + \frac{1}{f} \sum_i \epsilon_i {\psi_i'}^2 \ , \\
\dot X \cdot X' &=& f \gamma v x' + \frac{1}{f} \sum_i \epsilon_i (\omega_i + \phi_i) \psi_i' \ .
\end{eqnarray}

Recall that the canonical momentum densities associated to the string are
\begin{eqnarray}
\pi_{\mu}^0 &= &
-\frac{1}{2\pi \alpha'} g_{\mu\nu} \frac{(\dot X \cdot X') (X^\nu)' - (X')^2 (\dot X^\nu)}{\mathcal L} \ , \\
\pi_{\mu}^1 &=&
 -\frac{1}{2\pi\alpha'} g_{\mu\nu}  \frac{(\dot X \cdot X') (\dot X^\nu)- (\dot X)^2 ( X^\nu)'}{\mathcal L} \ ,
\end{eqnarray}
%
%
in terms of which the equations of motion for $x$ and $\psi_i$ can be written in the form
\[
\partial_\tau \pi_x^0 + \partial_\sigma \pi_x^1 = 0 \ ; \; \; \;
\partial_\tau \pi_{\psi_i}^0 + \partial_\sigma \pi_{\psi_i}^1 = 0 \ .
\]
From these equations and the stationary assumption,
it then follows that $\pi_x^1$ and $\pi_{\psi_i}^1$,
\begin{eqnarray}
\label{Peq}
\pi_x^1 &=& -\frac{1}{2\pi\alpha'} \frac{\partial {\mathcal L}}{\partial x'} \ , \\
\pi_{\psi_i}^1 &=& -\frac{1}{2\pi\alpha'} \frac{\partial {\mathcal L}}{\partial \psi_i'} \ ,
\label{Leq}
\end{eqnarray}
must be constant. A little bit of algebra also yields the relation
\be
\pi_t^1 = -\pi_x^1 v - \sum_i \pi_{\psi_i}^1\omega_i \ ,
\label{energyloss}
\ee
which can be thought of as an expression of energy
conservation: the power
being supplied through the endpoint of the string equals the force times the velocity.

Using the fact that $\pi_x^1$ and $\pi_{\psi_i}^1$ are constant, we can now in principle
solve for $x'$ and ${\psi_i}'$, after which
the expressions for $x'$ and ${\psi_i}'$ can
be integrated to yield the string profile.  In practice, we however need to do some
analysis to determine when the system of equations has real roots that
allow a string to exist for all $r$, $r_h < r < \infty$, and therefore leave this question aside for now.

In general, we would like to regulate the behavior of our spinning string by
ending it not at the boundary of AdS but at a D-brane at some $r_0$ large enough
such that the string can be considered a classical rather than a quantum
object in this curved background.  In practice, the precise definition of
large enough depends on the
details of the metric and string embedding,
but for the spinning D3-brane metric
(\ref{D3metric}) a sufficient condition is that $r_0$ be much larger
than the other scales in the problem, namely $r_h$ and $\ell_i$, or in field
theory language that the mass of the quark be large compared to the temperature
and the chemical potentials.
(See \cite{Herzog:2006gh} for a more careful analysis of this condition in the
case of zero chemical potential.)

There are several different natural boundary conditions to apply to these strings.
\begin{enumerate}
\item
We attach the string to a D-brane that is not spinning and that does not
extend in the $\psi_i$ directions, for which we have introduced a chemical
potential.  In this case, we should set
$\omega_i=0$.  These boundary conditions are compatible with the
equatorial strings of Ref.~\cite{Caceres:2006dj}.
\item
We attach the string to a D-brane that extends in the $\psi_i$ directions.
The string should feel no force from the D-brane, and we apply Neumann
boundary conditions.  We will see that these boundary conditions are
compatible with the uplifted strings of Ref.~\cite{Herzog:2006se}.
\item
The D-brane is spinning or has an electric field that causes the string
to rotate at some fixed rate $\omega_i$. Such an electric field in the
$\psi_i$ directions should correspond to a gradient in the
R-charge chemical potential.
\end{enumerate}
We shall return to a discussion of these D-branes in Section \ref{sec:D7necks}.

From conservation of the worldsheet currents, $\partial_a \pi_\mu^a=0$, it follows that
the string will gain energy, momentum, and R-charge through an endpoint at the D-brane, $r=r_0$, at a rate
\be
\frac{d E}{dt} = \left. -\pi_t^1 \right|_{r=r_0}  \; ; \; \; \;
\frac{d p}{dt} = \left. \pi_x^1 \right|_{r=r_0} \; ; \; \; \;
\frac{d \calj_i}{dt} = \left. \pi_{\psi_i}^1 \right|_{r=r_0}  \ .
\ee
Note that the total energy gain has contributions from both the momentum and
R-charge currents through (\ref{energyloss}).
In cases (1) and (3), generically the D-brane will apply a force to the endpoint of
the string.  In case (2), the solutions satisfy a no force condition in the
$\psi_i$ directions; we call such strings no-torque strings.

To explore these different boundary conditions, we will consider two general classes
of solutions.  In the first class, we explore cases (1), (2), and (3) but only turn on
one chemical potential and only allow the string
to spin in the $\psi_i$ direction corresponding to the $\phi_i \neq 0$.
We call such strings single charge strings.
In the second class, we impose the 
no-torque condition, but allow rotation in all the $\psi_i$ directions.

\subsection{Single charge spinning strings}
\label{sec:simple}

Assuming that the string is spinning in only one of the $\psi_i$ directions, $\psi_1=\psi$,
and that $\phi_2=\phi_3=0$,
the equations (\ref{Peq}) and (\ref{Leq}) can be massaged into the form:
\begin{eqnarray}
\label{Pexpand}
(x')^2 a_1 + 2 x' \psi' b_1+  (\psi')^2 c_1 +P^2  \beta I_1 &=& 0 \ , \\
(x')^2 a_2 + 2 x' \psi' b_2+  (\psi')^2 c_2 + L^2 \beta I_1 &=& 0\ ,
\label{Lexpand}
\end{eqnarray}
where we have defined $P \equiv 2 \pi \alpha' \pi_x^1$ and $L \equiv 2 \pi \alpha' \pi_\psi^1$ and
\be
I_1 = f^2 (-\alpha + v^2 \gamma) + \epsilon(\phi + \omega)^2 \ .
\ee
The precise form of $a_i$, $b_i$ and $c_i$ is included in
Appendix \ref{app:formulae}.

We want to establish a necessary condition for having a real solution for $x'$ and
$\psi'$ for all $r$, $r_h < r < \infty$.
The following two quantities will be important in this discussion:
\begin{eqnarray}
I_P &=& P^2 - f^2 \alpha \gamma + \gamma \epsilon (\phi + \omega)^2 \ , \\
I_L &=& L^2 - \alpha \epsilon + v^2 \gamma \epsilon \ .
\end{eqnarray}
For the backgrounds of interest, $I_1$, $I_P$ and $I_L$ will all be negative for large
radius $r$, far from the horizon $r_h$, but will be positive for $r$ close to the horizon.
Let $I_1(r_1) = 0$, $I_P(r_P)=0$ and $I_L(r_L)=0$, where $r_L$, $r_P$ and $r_1$
are all larger than the horizon radius $r_h$.

At the point $r=r_P$, the equation (\ref{Pexpand})
reduces to
\[
\left.
(-P^2 + f^2 v^2 \gamma^2)
\left( (\psi')^2 \alpha \epsilon+ \frac{P^2 \beta}{\gamma} \right)\right|_{r=r_P}=0
\]
which has no real solution unless $P^2 = f^2 v^2 \gamma^2|_{r=r_P}$ which in turn implies
that $I_1(r_P)=0$.
Similarly, at the point $r=r_L$, the equation (\ref{Lexpand}) reduces to
\[
\left. (-f^2 L^2 + \epsilon^2 (\phi + \omega)^2) \left( (x')^2 \alpha \gamma+ \frac{L^2 \beta}{\epsilon} \right) \right|_{r=r_L}=0
\]
which has no real solution unless
$L^2  = \frac{\epsilon^2}{f^2} (\phi + \omega)^2 |_{r=r_L}$ which in turn
implies that $I_1(r_L)=0$.  As we only expect one real root for $I_1$, $I_L$, and $I_P$ in the range
$r_h < r < \infty$, we should take $r_L = r_P = r_1$.

This critical radius $r_c\equiv r_1=r_P=r_L$ has an interesting physical significance, as was
pointed out by Refs.~\cite{Gubserjet, TeaneySolanajet}.  From the point of view of a two-dimensional
observer on the world-sheet of the string, there is a black hole horizon at this radius.
The induced worldsheet metric component $G_{\tau\tau} = \dot X^2$ is related to $I_1$ through
$f I_1 = \dot X^2$.  Thus where $I_1$ vanishes, $G_{\tau\tau}$ also vanishes.
We will sometimes refer to this radius as the effective horizon radius.

This necessary condition for the existence of a string stretching from the horizon to the boundary
of the space time also tells us the forces required to keep the string stationary.  From the above, we can immediately read off
\begin{eqnarray}
\label{onechargeP}
P \equiv 2 \pi \alpha' \pi_x^1 &=& \pm f(r_1) \gamma(r_1) v \ , \\
L \equiv 2 \pi \alpha' \pi_\psi^1 &=& \pm \frac{\epsilon(r_1)}{ f(r_1)} (\phi(r_1) + \omega) \ .
\label{onechargeL}
\end{eqnarray}
We can now finally also solve (\ref{Pexpand}) and (\ref{Lexpand}) to obtain expressions for $x'$ and
$\psi'$.  The answers are
\begin{eqnarray}
(x')^2 &=&
\frac{\beta \epsilon}{\alpha \gamma}
\frac{\left(\pm L v \gamma (\phi + \omega) + P (-\alpha + v^2 \gamma) \right)^2}
{\gamma \epsilon( L(\phi+\omega) \pm Pv)^2 - L^2 f^2 \alpha \gamma - P^2 \alpha \epsilon
-\alpha \gamma \epsilon I_1} \ ,
\label{singlexp}
 \\
(\psi')^2 &=&
\frac{\beta \gamma}{\alpha \epsilon}
\frac{\left(\pm L \left( \epsilon(\phi+\omega)^2  - f^2 \alpha \right)+ P v \epsilon (\phi+\omega) \right)^2}
{\gamma \epsilon( L(\phi+\omega) \pm Pv)^2 - L^2 f^2 \alpha \gamma - P^2 \alpha \epsilon
-\alpha \gamma \epsilon I_1} \ .
\label{singlepsip}
\end{eqnarray}

The equations for the forces $\pi_x^1$ and $\pi_\psi^1$ and the profiles $x'$ and $\psi'$
of the single charge string require some remarks.  The first is that all of these equations
have a dependence on the polar angles $\theta_a$ of the five sphere, either directly through
the $f(\theta_a, r)$ and $\epsilon(\theta_a, r)$ or indirectly through the
$\theta_a$ dependence of  the location of the roots of $I_1$, $I_P$, and $I_L$.

For the spinning D3-brane background with only one chemical potential $\phi_1 \neq 0$,
the $\theta_a$ dependence is easily understood.
The functions $f$ and $\epsilon$ become independent of the second polar angle $\theta_2$.
Moreover, the equations of motion for the string are satisfied in this single charge case only
for $\theta_1 = \pi/2$ or $\theta_1 = 0$.

The case $\theta_1=0$ is somewhat degenerate and
corresponds to the polar strings of Ref.~\cite{Caceres:2006dj}.  Here,
the $\psi$ circle in which the string can spin shrinks to zero size.
The metric coefficient $\epsilon$ vanishes at this value of $\theta_1$ as do
the parameters
$L$ and $\omega$.
One sees from (\ref{Pexpand}) and (\ref{Lexpand}) that $\psi'$ drops out of the
equations of motion.

The equatorial case $\theta=\pi/2$ is more intricate.  The nonspinning
 case $\omega=0$
corresponds to the equatorial strings of Ref.~\cite{Caceres:2006dj} while
the no-torque case $L=0$, we argue in Section \ref{sec:fived}, is the uplift of the single
charge 5d strings of Ref.~\cite{Herzog:2006se}.

Another remark concerns the near horizon behavior, $r \approx r_h$
of the string profile.
For the spinning D3-brane metric (\ref{D3metric}), the near horizon behavior
of the metric coefficients is given by
\be
\alpha(r) = \alpha'(r_h) (r-r_h) + {\mathcal O}(r-r_h)^2 \; ; \; \; \;
\beta(r) = \frac{\beta_{-1}}{r-r_h} + {\mathcal O}(1) \ .
\label{abexp}
\ee
Expanding (\ref{singlexp}) and (\ref{singlepsip}) near the horizon yields
\be
(x')^2 = \frac{v^2 \beta}{\alpha} + {\mathcal O}(r-r_h)^{-1} \; ; \; \; \;
(\psi')^2 = \frac{(\omega + \phi)^2 \beta}{\alpha} + {\mathcal O}(r-r_h)^{-1} \ .
\ee
Thus, the string profile generically has a logarithmically long tail due to
the near horizon behavior of $x'$ and $\psi'$.  The logarithmic divergence
from $x$ was already noticed in Refs.~\cite{Herzog:2006gh, Gubserdrag}.
The logarithmic divergence in $\psi$ causes the string to wrap the black
hole horizon an infinite number of times.  Most of the wrappings happen
exponentially close to the horizon, hence the red outline of the black hole
in Figure \ref{stringcartoon}b.

\subsection{No-torque strings}

In this Section, we consider strings spinning in all three $\psi_i$ directions
such that the applied force in the angular directions vanishes.
For the force applied by the D-brane in the $\psi_i$ directions to vanish,
we must set $\pi_{\psi_i}^1 = 0$ at the endpoint of the string, and since
$\pi_{\psi_i}^1$ is actually constant, it follows that $\pi_{\psi_i}^1=0$ everywhere
along the string.
This condition gives us a linear relation between $\psi_i'$ and $x'$
\be
\frac{\psi_i'}{\phi_i + \omega_i} = \frac{ \gamma v }{-\alpha  + \gamma v^2} x' \ ,
\ee
which allows us to solve easily for $x'$. In analogy with Eq.~(\ref{singlexp}), we obtain
\be
(x')^2 = \frac{P^2 \beta}{\alpha \gamma}
\frac{(\alpha -v^2 \gamma)^2}{-\alpha \gamma I_1 - P^2 (\alpha-v^2 \gamma)} \ .
\label{notorqueprofile}
\ee

We may also repeat the analysis of the single charge solutions in the previous Section regarding the existence
of a real solution all the way from the horizon to the D-brane. Demanding that
$\alpha - v^2 \gamma$, $ (\omega_i + \phi_i)$, and $(2\pi \alpha' \pi_x^1)^2 - f^2 \alpha \gamma$
vanish at the same critical radius $r_c$, we find that in order to
maintain the stationary solution, we need to apply a force at the end of the string amounting to
\be
\pi_x^1 = \left. \pm \frac{1}{2\pi \alpha'} f v \gamma \right|_{r=r_c} \ .
\label{notorqueresult}
\ee


One immediate
issue regarding the no torque solutions is the possible polar angle dependence of $f$, for which
we in the case of the spinning D3-brane solution choose $f^2 =\Delta$.
The force needed to keep the string stationary appears to depend on the polar angle, leading
to an instability towards the configuration requiring the least force, for which
the equations of motion are satisfied.  As discussed in the single charge case above,
two of the special values of
the $\theta_a$, for which this happens, are 0 and $\pi/2$.
In the case $\ell_1 \neq 0$ but $\ell_2 =\ell_3=0$, the force is minimized
by taking $\theta_1 = \pi/2$, for which $f(\theta_1{=}\pi/2) = H_1^{-1/3}$.
In another simple case $\ell_1 = 0$ but $\ell_2 = \ell_3 \neq 0$,
the force is minimized by taking $\theta_1 = 0$ for which $f(\theta_1{=}0) = H_2^{-1/6}$.
For the equal charges case, $\ell_1 = \ell_2 = \ell_3$, all dependence on the polar angles vanishes.

\subsection{Charge, momentum and energy of the spinning strings}
\label{sec:charge}

In this Section we consider the charge, momentum, and energy densities of our strings
in various limits in order to try to gain a better understanding of the field theory implications
of our solutions.
We first consider the momentum and R-charge density of the single charge spinning strings, and note that
from (\ref{singlexp}) and (\ref{singlepsip}), we find that
\begin{eqnarray}
(2 \pi \alpha' \pi_x^0)^2 &=&
\frac{\beta \gamma \epsilon}{\alpha}
\frac
{\left((P^2 - \alpha \gamma f^2 )v  \pm PL (\phi+\omega) \right)^2}
{\gamma \epsilon( L(\phi+\omega) \pm Pv)^2 - L^2 f^2 \alpha \gamma - P^2 \alpha \epsilon
-\alpha \gamma \epsilon I_1} \ ,
\label{pixzero} \\
(2 \pi \alpha' \pi_\psi^0)^2 &=&
\frac{\beta \gamma \epsilon}{\alpha}
\frac{\left( (L^2 - \alpha \epsilon) (\phi+\omega) \pm PLv \right)^2 }
{\gamma \epsilon( L(\phi+\omega) \pm Pv)^2 - L^2 f^2 \alpha \gamma - P^2 \alpha \epsilon
-\alpha \gamma \epsilon I_1} \ .
\label{pipsizero}
\end{eqnarray}

We would like to analyze these momentum and charge density expressions to see if the total
momentum and charge density of the string is finite.
At the horizon $r=r_h$, these densities typically diverge.
Recall that the near horizon behavior of the metric coefficients $\alpha$
and $\beta$ is given by (\ref{abexp}).
Expanding (\ref{pixzero}) and (\ref{pipsizero}) near the horizon, we find that
\[
(2 \pi \alpha' \pi_x^0)^2 = \frac{P^2 \beta}{\alpha} + {\mathcal O}(r-r_h)^{-1} \; ; \; \; \;
(2 \pi \alpha' \pi_\psi^0)^2 = \frac{L^2 \beta}{\alpha} + {\mathcal O}(r-r_h)^{-1} \ .
\]
Thus, the total momentum is finite if $P=0$ and the total
R-charge is finite if $L=0$.  Otherwise, both will be logarithmically
divergent because of the growth in the corresponding densities near the horizon.
Through (\ref{energyloss}), these divergences in the charge and momentum will
typically lead (for nonzero $\omega$ and $v$) to a divergence in the total energy
as well.

The divergences in the energy can be understood
physically.  As was pointed out in Ref.~\cite{Herzog:2006gh}, if we exert a
force $\pi_x^1$ in the $x$ direction for an infinite amount of time to keep the quark
moving at finite velocity, we will have done an infinite amount of work
which is now stored in the string.  Similarly, if a force $\pi_\psi^1$ is required to
keep the quark moving in the $\psi$ direction, we expect a similarly
divergent contribution to the total energy.

The equatorial case analyzed in Ref.~\cite{Caceres:2006dj}, where
$\pi_{\psi}^1 \neq 0$ and $\omega = 0$, will not produce a divergent energy contribution
because the torque does no work.  However, the string will still have a divergent
total R-charge.  Note that because the R-charge is associated to a global rather
than a gauged symmetry
in the field theory, there is no Coulombic interaction between R-charges and no
electrostatic contribution to the energy from having an infinite amount of R-charge
concentrated on a single string.
(The polar case analyzed in Ref.~\cite{Caceres:2006dj} in contrast is uncharged with respect
to the nonzero chemical potential.)

We next consider the momentum and R-charge density of the torqueless strings.
Based on the single charge results, we expect to find that the
total momentum is divergent but that the R-charge is not.
We easily obtain the results
\ba
{\mathcal L} &=& \sqrt{\fr{\alpha\beta\gamma I_1^2}{(-\alpha+v^2\gamma)P^2-\alpha\gamma I_1}}, \label{Lag}\\
2 \pi \alpha' \pi_{\psi_i}^0 &=&  \epsilon_i \sqrt{\fr{\alpha\beta\gamma(\phi_i+\omega_i)^2}{(-\alpha+v^2\gamma)P^2-\alpha\gamma I_1}}, \label{pipsi0}\\
2 \pi \alpha' \pi_{x}^0 &=& v\sqrt{\fr{\beta\gamma(P^2-\alpha\gamma f^2)^2}{\alpha((-\alpha+v^2\gamma)P^2-\alpha\gamma I_1)}} \label{pix0} \ ,
\ea
where we have used a modified definition of $I_1$, namely:
\[
I_1 = f^2 (-\alpha+v^2\gamma) + \sum_{i=1}^3 \epsilon_i (\phi_i+\omega_i)^2 \ .
\]
At the horizon $r=r_h$, one of these densities does in fact diverge.
We see that $\pi_{\psi_i}^0$ is finite at $r=r_h$ while $\pi_x^0$ diverges as
$1/(r-r_h)$.
We give more explicit expressions for these densites in the single-charge case,
three equal charges case, and the equatorial strings of Ref.~\cite{Caceres:2006dj}
in Appendix \ref{app:specialcases}.

On the field theory side, we expect these divergent charge, momentum, and energy densities
to be cutoff by physical processes.  For example, as was discussed in Ref.~\cite{Herzog:2006gh},
one can think of a process in which a quark-antiquark pair is separated by an electric field.
After a sufficiently long
time, the single quark solution cut off at some radius $r \gtrsim r_h$ becomes half
the quark-antiquark solution.  There is also a $1/N$ suppressed process in which the charge
and energy of the quark leaks into the surrounding medium.  On the gravity side,
the
leakage is described by the back reaction of the string on the black hole geometry
and was analyzed in the zero chemical potential case in Refs.~\cite{Gubserwake,Paul}.

Now we turn to the task of showing that these torqueless strings
are the uplift of the 5d strings considered in Ref.~\cite{Herzog:2006se}.

\section{Comparison with 5d strings}
\label{sec:fived}

In Ref.~\cite{Herzog:2006se}, dragging strings in a generic, asymptotically AdS geometry were studied.
Given some assumptions, we can lift those strings to the spinning strings studied here.
First, let us recall these dragging string solutions.  We assume a metric of the form
\be
ds^2  = -\alpha(r) dt^2 + \beta(r) dr^2 + \gamma(r) d x^2 \ .
\ee
As $r \to \infty$, the metric should approach that of $AdS_3$ with a radius of curvature $R$:
\be
\alpha \to R^2 r^2 \; ; \; \; \;
\beta \to \frac{R^2}{r^2} \; ; \; \; \;
\gamma \to R^2 r^2 \ .
\ee
The coefficients $\alpha$, $\beta$, and $\gamma$ here are essentially the same
as in (\ref{genericmetric}).
The space is assumed to have a horizon at $r=r_h$ with $\alpha$ and $\beta$
of the form (\ref{abexp}).

We assume the string has a time dependence of the form $\dot x = v$, for which the equation of
motion reduces to
\be
(x')^2 = \frac{\beta P^2 (-\alpha + \gamma v^2)}
{\gamma \alpha(-\gamma \alpha + P^2)} \ ,
\label{fivedprofile}
\ee
where again $P = 2 \pi \alpha' \pi_x^1$.
From the asymptotic form of $\alpha$ and the other metric components, it is clear
that the numerator and denominator of this expression for $(x')^2$ will be negative
for large $r$ and positive for $r \approx r_h$.  To have a string that stretches all the
way from $r=r_h$ to the boundary, we must require that the numerator and denominator
have a zero in the same place.  This condition allows us to solve for $\pi_x^1$:
\be
\pi_x^1 = \pm \frac{1}{2 \pi \alpha'} v \gamma(r_c)
\ee
where $-\alpha(r_c) + \gamma(r_c) v^2 = 0$.
We make two important points:
\begin{enumerate}
\item
The location of the critical radius $r_c$ is the same for both the no-torque strings 
and these dragging strings.
\item
The force $\pi_x^1$ required to maintain the speed of the no-torque strings  (\ref{notorqueresult})
and the dragging
strings is the same provided that $f(r_c) = 1$.
\end{enumerate}

At the end of Ref.~\cite{Herzog:2006se}, a specific example of these dragging strings was analyzed,
strings in the dimensionally reduced spinning black D3-brane background.  The five dimensional
metric (\ref{fivered}) was used for the calculation.  Thus, the choice $f^2 = \Delta$ was made.
Three cases were considered: (1) $\ell_1 = \ell_2 = \ell_3$; (2) $\ell_1=0$ and $\ell_2=\ell_3 \neq 0$;
and (3) $\ell_1 \neq 0$ and  $\ell_2 = \ell_3=0$.
In case (1), for the dimensionally reduced
metric employed, $f(r_c)=1$ and the 5d strings require the same amount
of force to be dragged as the no-torque strings.
In the other two cases $f(r_c) \neq 1$, and the strings do not uplift properly.

However,
it is a choice to set $f^2 = \Delta$.  We could absorb $\Delta$
into the five dimensional metric and choose $f=1$.
More subtly, if we wanted to maintain the independence of the reduced metric on
the auxiliary angular coordinates $\theta_a$, then
in case (2), we could define a new $f^2$ such that
\be
f^2_{(2)} = H \mu_1^2 + \mu_2^2 + \mu_3^3  = H^{1/3} \Delta\ .
\ee
whereas for case (3), we choose
\be
f^2_{(3)} = \mu_1^2 + H (\mu_2^2 + \mu_3^2) = H^{2/3} \Delta \ ,
\ee
with the corresponding now angle independent shifts
in $\alpha$, $\beta$, $\gamma$, and $\epsilon_i$.

Morally, this redefinition of $f$ should be analogous to the relation between string frame
and Einstein frame in ten dimensional supergravity theories.  In the cases where
the R-charge chemical potentials are not all equal, scalars in the five dimensional
effective supergravity description of the black hole have nontrivial expectation value.
In a sense we have not been able to make precise, these nontrivial expectation values
should allow one to describe an effective five dimensional string frame metric where $f=1$.

Both the 5d strings of Ref.~\cite{Herzog:2006se} and the no-torque
strings discussed here have an effective horizon at the same radius $r=r_c$, and, if
we set $f(r_c)=1$, require
the same amount of force to be dragged, but the two strings will have different profiles.
The equations for $x'$ in the two cases are different.  For example, as can
be seen from (\ref{notorqueprofile}), $x'$ for the no-torque strings depends on $I_1$ which in turn
depends on metric coefficients in the $S^5$ directions.
In contrast, the 5d $x'$ by definition can depend only on the metric coefficients
$\alpha$, $\beta$, and $\gamma$.  It would be interesting to see if this difference in profile
has any observable consequences in the field theory, for example in the energy density
of the quark wake.

\section{D7-brane boundaries}
\label{sec:D7necks}

In this Section, for simplicity we will assume the string can only spin in one of the $\psi$
directions.

To avoid having infinitely massive strings, we can have the strings stretch from the horizon
to a D7-brane located at some large radius $r=r_0$.  At this D7-brane, the boundary condition
must be consistent with the type of spinning strings.
For example, for the torqueless strings,
the endpoint of the string must be free to spin around the $\psi$ direction
as depicted in Figure \ref{stringcartoonfancy}.\footnote{%
 Another possibility is that the D-brane be allowed to rotate with the
 string.  It would be interesting to see if such spinning D-branes could
 be constructed.
}
  In contrast,
for the equatorial strings of Ref.~\cite{Caceres:2006dj},
the D-brane shown in Figures \ref{stringcartoon}b and \ref{stringcartoon}c
must exert a force on the string both in the
$\psi$ direction to keep the string from spinning and in the $x$ direction to keep the string
moving at a constant $v$.
In the degenerate polar string case of Ref.~\cite{Caceres:2006dj},
the quark is uncharged with respect to the nonzero chemical potential
and the relevant $\psi$ circle is vanishingly small.

The D7-brane appropriate for the equatorial strings of Ref.~\cite{Caceres:2006dj}
was considered for other reasons in Ref.~\cite{Clifford}
and is a generalization of the D7-brane considered by
Ref.~\cite{KatzKarch} in pure AdS.  The D7-brane wraps a variable size $S^3$ inside
the transverse $S^5$ that shrinks to zero at the D7-brane's point of closest approach
to the stack of D3-branes. It is at this point of closest approach $r=r_0$ that it is
natural to attach the string to minimize the string's energy.  At this point,
the $\psi$ angle of the string is fixed.  As we work in a large $N$ limit where the D7-brane
is infinitely more massive than the string, the D7-brane will not respond
to the force from the string in the $\psi$ direction.

The D7-brane for the polar string of Ref.~\cite{Caceres:2006dj} has not to 
our knowledge appeared
in the literature, but it involves nothing qualitatively new.  It is morally just
a rotation of the D7-brane appropriate for the equatorial string although
its energetics and profile will be quantitatively different because of the 
squashing of the $S^5$ in the black hole geometry.

The D7-brane for the torqueless string requires something new.  We consider a pair of parallel
D7-branes wrapping variable size $S^3$'s.  Now at the point of closest approach
to the stack of D3-branes, the D7-branes develop a neck and merge into each other, which we illustrate
in Fig.~\ref{stringcartoonfancy}.
The strings can spin freely inside this neck.  This configuration is reminiscent
of the D8-D4 system considered by Sakai and Sugimoto \cite{SakaiSugimoto}.
Note that the D7-brane configuration is uncharged with respect to the nonzero chemical potential
while the string, according to (\ref{pipsizero}), will have finite R-charge.
One way of understanding why the string here does not have infinite R-charge is
that the D7-branes, being uncharged, have no means of supplying it.

To explain these D7-branes, we work first in pure $AdS_5 \times S^5$ and then
discuss briefly the generalization to the spinning black D3-brane background.
Recall the metric in this pure case is
\be
ds^2 = h(r)^{-1/2} (-dt^2 + d \vec x^2) + h(r)^{1/2} \delta_{ij} dy^i dy^j
\ee
where $i = 1, \ldots, 6$, $r^2 = \sum_i (y^i)^2$, and $h(r) = R^4/r^4$.
As an ansatz, we take the D7-brane to fill the Minkowski space directions
and four of the directions in the transverse $\mathbb{R}^6$.  As a result,
the D7-brane will not feel the warp factor, and it is clear that
a simple embedding that solves the equations of motion is
to take two of the $y^i$ constant, leading to a D7-brane that wraps
a vanishing $S^3$ at the point of closest approach to the origin.
We rewrite the metric on $\mathbb{R}^6$ as
\be
ds_{\mathbb{R}^6}^2 = dr^2 + r^2(d\theta^2 + \sin^2 \theta d \psi^2 + \cos^2 \theta d \Omega_3^2)
\ee
where $d\Omega_3^2$ is the line element on a unit $S^3$.  In these coordinates,
taking two $y^i$ constant amounts to setting $\psi$ and $r \sin \theta$ constant.  Indeed,
when $r$ is smallest, $\theta=\pi/2$ and the $S^3$ shrinks away.

The solution with a neck can be obtained as follows.
Let $\chi = r \cos \theta$ and
$\xi = r \sin \theta$.  Assuming that $\psi=\mbox{const}$, the Lagrangian for the
embedding $\xi(\chi)$ is
\[
{\mathcal L} = - \chi^{3} ((\xi')^2 + 1)^{1/2}
\]
leading to the solution
\[
\xi = \int_C^\chi \frac{C^3 dx}{\sqrt{x^6 - C^6}} + \xi_0\ .
\]
For $C=0$, we recover the $y^i = \mbox{const}$ solution.
However, for $C \neq 0$,
we get a pair of D7-branes joined by a neck, inside of which a string might spin.
This curve $\xi(\chi)$ gives the profile of one of the D7-branes.  The other D7-brane
joins smoothly onto this curve, and its profile is obtained by flipping the sign of the integral.

There are two masses and a chiral condensate associated with this necked D-brane
configuration.  We can read off these numbers from the asymptotic
expansion of $\xi$ at large $\chi$,
\[
\xi = a C + \xi_0 + \frac{C^3}{2 \chi^2} +  \ldots
\]
where
\[
a \equiv
\int_1^\infty \frac{dx}{ \sqrt{x^6-1}}= - \frac{\sqrt{\pi} \Gamma(1/3)}{\Gamma(-1/6)}=  0.70109\ldots \ .
\]
The first term in the expansion gives the mass of the hypermultiplet corresponding
to one of the D7-branes, $m_+ \sim a C + \xi_0$.
The other mass is $m_- \sim -a C + \xi_0$.  The second term in the
expansion is proportional to the chiral condensate, $\langle q \bar q \rangle \sim C^3$.

To see how the string can spin inside this D7-brane pair, note that
inside the neck, if we set $\xi_0=0$, then when $\chi = C$, the polar angle $\theta=0$ vanishes
and the
$S^3$ wrapped by the D7-brane pair is large.  The string is free to spin inside this $S^3$.

Unfortunately, this solution is higher in energy than the pair of D7-branes without
such a neck.  Indeed, reinterpreting ${\mathcal L}$ as an energy density for our static
configuration,
\[
\Delta V = \lim_{\Lambda \to \infty} \left(
\int_C^\Lambda \frac{\chi^6}{\sqrt{\chi^6-C^6}} d\chi - \int_0^\Lambda \chi^3 d \chi \right) =
\frac{C^4 \sqrt{\pi} \Gamma(-2/3)}{6 \Gamma(-1/6)}
> 0 \ .
\]

By continuity, we expect that this D7-brane pair with neck configuration will continue
to exist for small temperature $T$ and chemical potential $\Phi$.
Also by continuity, given that the D7-brane pair with neck is unstable at the origin of the
$\Phi T$ plane, we expect the D7-brane pair with neck will
remain unstable in a small region around the origin.
The full stability analysis will need to be done numerically, and we leave such
an analysis for the future.

\section{Stability analysis of spinning strings}
\label{sec:stability}

In this Section, we begin an analysis of the stability of the single charge string
described by Eqs.~(\ref{singlexp})--(\ref{singlepsip}) with respect
to small fluctuations.  The main purpose of this paper is to
 explain the discrepancy between the strings of Refs.~\cite{Herzog:2006se} and
\cite{Caceres:2006dj}, and a full stability analysis is beyond our scope.  However,
the fact that the fluctuations we consider are stable gives us confidence that
both the equatorial strings of Ref.~\cite{Caceres:2006dj} and the no-torque strings of
Ref.~\cite{Herzog:2006se}
have meaningful field theory counterparts in ${\mathcal N}=4$ SYM at
nonzero temperature and R-charge chemical potential.  Moreover,
these fluctuations may be useful eventually in calculating momentum and R-charge
two-point correlation functions for the corresponding heavy quarks
 \`a la Refs.~\cite{Gubserjet} and \cite{TeaneySolanajet}.\footnote{%
 Similar stability analyses have been carried out before for mesonic solutions
 where both ends of the string touch the D7-brane \cite{stability}.
 The stability analysis here is slightly more involved because the 
 effective horizon on the string makes the problem non-Hermitian.
 }

We will consider fluctuations only in the directions $x$ and $\psi$ in which
the string is moving, and in a direction we call $\theta$, which may be taken
to be, for example, $\theta_1$ of the spinning D3-brane metric (\ref{D3metric}).
Because of the $\theta$ fluctuations,
we need a slight generalization of our generic metric (\ref{genericmetric}),
\be
ds^2 = f \left[ - \alpha \, dt^2 + \beta \, dr^2 + \gamma \, d  x^2 \right] +
\eta \, d\theta^2 +
\frac{\epsilon}{f}(d\psi + \phi \, dt)^2 \ .
\ee

To obtain an effective Lagrangian for the fluctuations, we expand ${\mathcal L}$
(\ref{calLdef}) 
to second order in $\delta x$, $\delta \psi$ and $\delta \theta$.  The first order terms
will vanish because we evaluate ${\mathcal L}$ at a solution to the equations of
motion.  Using Eqs.~(\ref{singlexp})--(\ref{singlepsip}),
the resulting second order contribution is
\begin{eqnarray}
 \delta^2 {\mathcal L} &=&
 - \frac{1}{2 {\mathcal L}_0} \left(
 I_L (\delta \psi')^2 +
 2 (LP - v \gamma \epsilon (\phi+\omega)) \delta \psi' \delta x' +
 I_P (\delta x')^2 +
 I_1 \frac{\eta}{f} (\delta \theta')^2
 \right) \nonumber \\
 &+&
  \frac{{\mathcal L}_0 f^2 }{2 I_1^2}
\left(
I_L (\delta \dot \psi)^2 +
2 (LP - v \gamma \epsilon (\phi+\omega) ) \delta \dot \psi \delta \dot x +
I_P (\delta \dot x)^2
+ I_1 \frac{\eta}{f} (\delta \dot \theta)^2
\right) \nonumber \\
&+&\frac{ -f^2 L v \gamma + P \epsilon (\phi+\omega)}{I_1}
(\delta \psi' \delta \dot x - \delta \dot \psi \delta x') \nonumber \\
&+&
\frac{{\mathcal L}_0}{2\epsilon I_1} \left(
 I_L f (\partial_\theta^2 f) + \frac{1}{2 \epsilon} (\partial_\theta^2 \epsilon)
(-f^2 L^2 + \epsilon^2 (\phi+\omega)^2 )
\right) (\delta \theta)^2\; ,
\label{masterfluct}
 \end{eqnarray}
 where
 \be
 {\mathcal L}_0 =
 \frac{-\sqrt{\alpha \beta \gamma \epsilon} I_1}
 {\sqrt{\gamma \epsilon( L(\phi+\omega) \pm Pv)^2 - L^2 f^2 \alpha \gamma - P^2 \alpha \epsilon
-\alpha \gamma \epsilon I_1}} \ .
 \ee
 The time derivatives, denoted by a $\dot {}$, are with respect to a new time
 $t \to t + F(r)$ where $F'(r) = (X' {\cdot} \dot X) / \dot X^2$.  This redefinition
 eliminates cross-terms of the form $\delta \dot x \, \delta x'$ and makes manifest
 the effective horizon on the string worldsheet at $r=r_c$.
 A similar redefinition was made in Refs.~\cite{Gubserjet} and \cite{TeaneySolanajet}.
Note that in general the $\delta x$ and $\delta \psi$ fluctuations are coupled.
Also, there is an implicit assumption here that we evaluate the Lagrangian
at $\theta=\pi/2$, where the equation of motion for $\theta$ is satisfied
for the spinning D3-brane metric (\ref{D3metric}).\footnote{%
  Instead of interpreting 
$\delta \theta$ as a fluctuation in the polar angle of the string, one may also interpret
$\delta \theta$ as a fluctuation in the gauge theory directions perpendicular to the direction
of motion.  In this case, the $(\delta \theta)^2$ term in the action must be taken to vanish
and $\eta$ reinterpreted as the $g_{xx}$ metric component.
}

\subsection{Equatorial string fluctuations}

We now specialize to the equatorial solution of Ref.~\cite{Caceres:2006dj}.
Using the fact that $\omega=0$ and plugging in the relevant values of $P$ and $L$
from Eq.~(\ref{CGPLvalues}), one finds
\begin{eqnarray}
\delta^2 {\mathcal L} &=& \frac{1}{\sqrt{1-v^2}}
\Bigg[
\frac{1}{\dot X_0^2} \left(
\frac{r^2}{2R^2} (\delta\dot x)^2 + \frac{R^2}{2}(1-v^2) ((\delta \dot \psi)^2 + (\delta \dot \theta)^2) \right)
\nn
&&
- \frac{\dot X_0^2}{1-v^2} \left( \frac{r^2}{2 R^2} (\delta x')^2 + \frac{R^2}{2} (1-v^2)
\left( (\delta\psi')^2 + (\delta\theta')^2 \right) \right) \Bigg],
\end{eqnarray}
where
\be
\dot X_0^2 = \frac{r_h^2(r_h^2 + \ell^2) - r^4 (1-v^2)}{r^2 R^2}
\ee
and, quite remarkably, all cross terms have vanished.

The fluctuations in the $\psi$ and $\theta$ directions are 
simple to analyze and reveal no instabilities.
As we can see from the fluctuation Lagrangian, the equations of motion
for $\delta \psi$ and $\delta \theta$ are identical.
Moreover, these fluctuation equations
can be solved by plane waves.  By defining a new radial
coordinate $r_*$ such that $dr_* = -dr / \dot X_0^2$, we find that $\delta \psi \sim \delta \theta
\sim e^{i k r_* - i \omega t}$, where the wave vector satisfies the modified dispersion relation
$k^2 = (1-v^2) \omega^2$.  In terms of this new radial coordinate, the position of the
effective horizon of the string $r=r_1$ becomes $r_* \to -\infty$.

The boundary conditions are
different for the fluctuations in these two angular directions.
We take this single charge string to end on the D7-brane of Ref.~\cite{Clifford}
described in the Introduction and Section \ref{sec:D7necks},
a cartoon of which is drawn in Figures \ref{stringcartoon}b and c.
For the $\psi$ fluctuations, we take Dirichlet boundary conditions at the
D7-brane because the D7-brane lies at a constant value of $\psi$.
In the $r\theta$ plane, the brane on the other hand lies along a curve $r(\theta)$; at the point where the string
attaches to the D7-brane, by symmetry the derivative of $r'(\theta)$ vanishes and to
quadratic order in the fluctuations, we may take the boundary conditions to be Neumann
in the $\theta$ direction.
At the effective horizon of the string $r_* \to -\infty$,
we take causal boundary conditions where
the waves pass into the horizon.  Given the plane wave nature of the solutions and these
relatively tame boundary conditions, there can be no instability in the $\theta$
or $\psi$ directions.

The fluctuations in
 $\delta x$ were already studied in Ref.~\cite{Herzog:2006gh}.
Taking a time dependence of the form $e^{-i \omega t}$,
we rewrite the equation of motion for $\delta x$ as
\be
 \left( 1- y^{-4} \right) \partial_y \left( y^4 \left( 1-y^{-4} \right) \partial_y \delta x \right)
+ w^2 \delta x = 0 \ ,
\label{oldeq}
\ee
where we have introduced $w^2 \equiv R^4 \omega^2  / [r_1^2 (1-v^2)]$ and
$y \equiv r/r_1$.
We again impose ingoing boundary conditions at the effective horizon $r=r_1$
and Neumann boundary conditions at the D7-brane $r=r_0$.
Precisely this same differential equation governs the fluctuations of the
straight $v=0$ string at zero chemical potential and was studied in detail in
Section 3 of Ref.~\cite{Herzog:2006gh} where no instabilities were found.  The smallest
quasinormal mode is purely imaginary and lies in the lower half of the complex
$\omega$ plane.    With our assumed time dependence, quasinormal modes
in the lower half of the complex plane cannot lead to instabilities.
The higher quasinormal modes will typically have even more negative
imaginary parts, as we have indeed been able to observe numerically.
It turns out fluctuations of the string in the gauge theory directions orthogonal to the 
direction of motion
are also governed by (\ref{oldeq}).  Thus we expect no instabilities for these
fluctuations either.

Using the results in Appendix \ref{app:rigorous}, we can be more rigorous.
We define a new radial variable $r_*$ such that $dy / dr_* = y^2 (1-y^{-4})$
and we rescale the fluctuations $ y \, \delta x(y) =  \delta \chi(r_*)$.
With these changes, the equation (\ref{oldeq}) becomes 
a Schr\"odinger problem
\be
\left(- \partial_{r_*}^2 + 2 y^{-6} (y^8-1) \right) \delta \chi = w^2 \delta \chi \ .
\ee
The potential function for this Schr\"odinger problem vanishes 
at $y=1$ or equivalently at $r_* \to -\infty$.
Given the rescaling by $y$, the outgoing boundary conditions
for $\delta x(y)$ remain outgoing boundary conditions for $\delta \chi(r_*)$.  However,
the Neumann boundary conditions for $\delta x$ become more complicated
mixed boundary conditions for $\delta \chi$, namely $\delta \chi'(r_*) = \delta \chi(r_*)\, \mbox{const} $,
but which are still self-adjoint.

In Appendix \ref{app:rigorous}, we give a proof that given these conditions on
the potential and boundary conditions on $\delta \chi$, stability amounts to proving
the absence of bound states for the potential $V(r_*)$.  Given the monotonic
increasing nature of $V(r_*)$, there are clearly no such bound states and the fluctuations are
stable.

\subsection{No-torque string fluctuations}

For the single charge no-torque solution of Eqs.~(\ref{singlexp})--(\ref{singlepsip}), it is a relatively simple task to rewrite Eq.~(\ref{masterfluct}) in the form
\ba
\delta^2 {\mathcal L} &=&-\fr{1}{2{\mathcal L}_0}\fr{r_1^2+\ell^2}{r^2(r_1^2+\ell^2)+r_1^4}\Bigg[\fr{r^2(r^2(r_1^2+\ell^2)+r_1^2(r_1^2+v^2 \ell^2))}{(1-v^2)(r_1^2+\ell^2)R^2}
\bigg\{\dot X_0^2\(\delta x'\)^2-\fr{{\mathcal L}_0^2}{\dot X_0^2}(\delta\dot x)^2\bigg\}\nn
&+&R^2(r^2+r_1^2+\ell^2)\bigg\{\dot X_0^2\(\delta \psi'\)^2-\fr{{\mathcal L}_0^2}{\dot X_0^2}(\delta\dot \psi)^2\bigg\}
+\fr{2v\omega R^2 r^2}{1-v^2}
\bigg\{\dot X_0^2\delta x' \delta\psi'-\fr{{\mathcal L}_0^2}{\dot X_0^2}\delta\dot x\delta\dot\psi \bigg\}\Bigg]\nn
&+&\fr{v \omega r_1^2 (r^2-r_1^2)}{r^2\dot X_0^2}\bigg\{\delta\dot x\delta\psi'-\delta x'\delta\dot\psi\bigg\}
\nn
&-& \frac{R^2}{2} \left\{ \fr{\dot X_0^2}{{\mathcal L}_0} \(\delta\theta'\)^2
- \fr{{\mathcal L}_0}{\dot X_0^2} (\delta\dot\theta)^2 \right\}
- \frac{1}{2} {\mathcal L}_0 V (\delta \theta)^2 \ .
%
\ea
Here, the various functions have the forms
\ba
\omega
&=&
-\fr{\ell r_1}{R^2}\sqrt{\fr{1-v^2}{r_1^2+\ell^2}} \; , \; \; \;
V = -\fr{\ell^2(2r_1^2+\ell^2)}{(r_1^2+\ell^2)r^2+r_1^4} \ ,
\\
{\mathcal L}_0&=&\fr{\sqrt{1-v^2}\((r_1^2+\ell^2)r^2+r_1^4\)}{\sqrt{(r_1^2+\ell^2)((r_1^2+\ell^2)(r^2+r_1^2)^2+\ell^4r^2+v^2r_1^2 \ell^2(r_1^2+\ell^2))}} \ ,\\
\dot X_0^2 &=&-\fr{(1-v^2)(r^2-r_1^2)((r_1^2+\ell^2)r^2+r_1^4)}{R^2(r_1^2+\ell^2)r^2}\ . 
\ea

We analyze fluctuations in the $\theta$ direction.  The equation of motion becomes
very simple to analyze after defining a new radial coordinate $r_*$ where
$dr_* = -{\mathcal L}_0 dr / \dot X_0^2$, as it can be written as
\be
-\frac{\partial^2}{\partial r_*^2} \delta \theta + \frac{\ell^2 (\ell^2 + 2 r_1^2) (1-v^2)}{r^2 R^2 (\ell^2 + r_1^2)}
\left( 1 - \frac{r_1^2}{r^2} \right) \delta \theta = -\frac{\partial^2}{\partial t^2} \delta \theta \ .
\label{thetafluct}
\ee
Assuming a time dependence of the form $\delta \theta \sim e^{-i \omega t}$, this partial
differential equation reduces to a Schr\"odinger problem with energy $E =  \omega^2$
and potential $V \sim r^{-2} (1-r_1^2/r^2)$.
Now the string is taken to end on the D7-brane pictured in Figure \ref{stringcartoonfancy}.
The boundary conditions are Neumann at the D7-brane
and ingoing at the effective horizon.
Although the eigenfunctions will no longer be
plane waves, from the bounded form of the potential and the boundary conditions, it is clear there
can be no instability from fluctuations in the $\theta$ direction.
Indeed our numerical analysis revealed quasinormal modes only in the lower half
of the complex $\omega$ plane.  

More rigorously, we can invoke the results of Appendix
\ref{app:rigorous}.  The potential $V$ is a bump function that goes to zero both
as $r \to r_1$ and as $r \to \infty$.  We cut off the potential at some large $r_0$ with 
a D7-brane which presents the possibility of having some long lived state that
is trapped between the D7-brane and the top of the bump.  However, since
the potential at the horizon $r=r_1$ is always lower than the potential at $r=r_0$, 
such a state can at most be metastable and cannot be a bound state.  The absence
of bound states of $V$ proves the fluctuations are stable.
The behavior of the fluctuations in the gauge theory directions orthogonal to the direction of 
motion are governed by a differential equation qualitatively similar to (\ref{thetafluct}).
The potential function appears to be a similar type of bump function, but we have not
analyzed it in full generality.

Although the fluctuations we have analyzed revealed no instabilities,
we have not done a complete stability analysis of all of our various strings.  In fact,
there are some glaring omissions.
We would like to be able to analyze the $\delta x$-$\delta \psi$ system of equations for
this no-torque string.
Thus far, we have found no simple approach to decoupling the equations,
but we hope to return to this case numerically in the future.  Fluctuations around
the polar string of Ref.~\cite{Caceres:2006dj} promise to be very interesting because
in this case the small fluctuations can rotate about the $\psi$ direction
and leave open the possibility
of super-radiant instabilities.

\section{Discussion}


At finite R-charge chemical potential, the drag force experienced by a heavy
probe quark in the ${\mathcal N}=4$ SYM plasma depends
on the details of the dual geometry transverse to the asymptotically
$AdS_5$ part.  The transverse
geometry at finite R-charge chemical potential is a squashed
$S^5$.    Depending on how the string is oriented in the squashed
$S^5$, for example whether it is a polar or an equatorial string, and depending
on how the string is moving in the $S^5$, for example whether it is spinning or
not, the drag force will be different.

Much of the dependence on the transverse geometry can be understood
from the field theory.  As was discussed in the Introduction,
the polar strings correspond
to single quarks in massive ${\mathcal N}=2$ hypermultiplets that are not
charged under the chemical potential, while the equatorial strings come
from ${\mathcal N}=2$ hypermultiplets that are.  The D7-branes
used to anchor the no-torque strings are unstable and their field
theory interpretation unclear, yet the geometry is reminiscent of
the Sakai-Sugimoto model \cite{SakaiSugimoto} where the field
theory has a broken chiral symmetry and a nonzero chiral condensate.

One could have hoped that fundamental heavy probes of nonabelian plasmas
would exhibit universal or quasi-universal behavior.  Such universality leaves
the door open for the exciting possibility that what we learn about ${\mathcal N}=4$
SYM in the strongly interacting limit might also hold true for QCD at temperatures
not too much above the deconfinement transition.
In light of such a possibility, the dependence of the drag
force on the $S^5$ is bad news.  Other gauge/gravity dualities will have
different transverse geometries and correspondingly different drag forces.

Despite the dependence on the $S^5$, the formula (\ref{onechargeP}) for the drag force
is in a peculiar sense universal.  The result states that the drag force is
\be
\pi_x^1 = \frac{1}{2\pi \alpha'} g_{xx}(r_1) v \ ,
\label{universaldrag}
\ee
where $g_{xx}(r_1)$ is the ${xx}$-metric component of the geometry evaluated
at the effective horizon of the string.  The entropy of a black hole
is proportional to its area, and in this translationally invariant setting, the entropy density
 would scale as $(g_{xx}(r_h))^{d-1}$ for a field theory in $d$ dimensions.
 We speculate that there is a sense in which $g_{xx}(r_1)$
 can be thought of as an entropy-like quantity associated to the probe heavy quark.
 Indeed, as was pointed out in Ref.~\cite{Herzog:2006se}, in the limit $v \to 0$, the black hole
 horizon and the effective string horizon coincide, $r_1 \to r_h$;  in this limit, $g_{xx}(r_1)$
 is related to the entropy density of the field theory itself.\footnote{%
  This formula (\ref{universaldrag}) has appeared before for 5d strings as Eq.~(17) of
  Ref.~\cite{guijosadrag}.
  A similar behavior for the drag force was observed in ${\mathcal N}=4$ SYM
  at nonzero baryon chemical potential \cite{obannontoappear}.
  }

\section*{Acknowledgments}
We would like to thank Paul Chesler, Andreas Karch, Andy O'Bannon,
Maurizio Piai, Toni Rebhan, Misha Stephanov, and Larry Yaffe for discussions.
C.~H. would like to thank the Perimeter Institute and the Benasque Center
for Science for hospitality where part of this work was completed.
The work was supported in part by the U.S.~Department of Energy under Grant No.~DE-FG02-96ER40956.
A.V.~was in addition supported by the Austrian Science Foundation, FWF, project no. M1006.

\appendix

\section{Spinning String Formulae}
\label{app:formulae}

The expressions $a_i$, $b_i$, and $c_i$ determining the two quadratic
forms of (\ref{Pexpand}) and \ref{Lexpand}) are as follows:
\begin{eqnarray*}
a_1 &=& \gamma (-\alpha f^2 + \epsilon(\phi+\omega)^2) I_P \ , \\
b_1 &=& -v \gamma \epsilon (\phi+\omega) I_P \ , \\
c_1 &=& \epsilon(P^2(-\alpha + v^2 \gamma) + \epsilon \gamma^2 v^2 (\phi+\omega)^2 ) \ ,\\
a_2 &=& \gamma (L^2 (-\alpha f^2 + \epsilon (\phi+\omega)^2) + \epsilon^2 \gamma v^2 (\phi+\omega)^2) \ , \\
b_2 &=& -v \gamma \epsilon (\phi + \omega) I_L\ ,  \\
c_2 &=& -\epsilon (\alpha -v^2 \gamma) I_L \ .
\end{eqnarray*}
The determinants of these quadratic forms are
\begin{eqnarray*}
a_1 c_1 - b_1^2 &=& -P^2 \alpha \gamma \epsilon I_1 I_P\ , \\
a_2 c_2 - b_2^2 &=& - L^2 \alpha \gamma \epsilon I_1 I_L \ .
\end{eqnarray*}
From the determinants, we see that the quadratic forms will have eigenvalues of opposite
sign when $r < r_{min} \equiv  \mbox{min}(r_1, r_P, r_L)$ and also when
 $r> r_{max} \equiv \mbox{max}(r_1, r_P, r_L)$.

\section{Special cases for the charge and momentum densities}
\label{app:specialcases}

\subsection{Single charge, no-torque case}
In the single charge case
where
$\ell_1\equiv \ell$, $\phi_1\equiv \phi$ and all other $\ell_i$'s and $\phi_i$'s vanish,
we can give explicit expressions for the charge and momentum densities and the total charge.
In this case, the minimal force condition implies $\theta_1=\pi/2$, and all other angles with the exception of $\psi_1$ decouple.
We can use the spinning D3-brane metric (\ref{D3metric}) to write these densities in their explicit forms.
Taking $f^2=\Delta$ as above, the effective horizon $r_1$
is straightforwardly solved from the condition $(\alpha-v^2\gamma)\mid_{r=r_1}=0$, producing
\be
2r_1^2 =  \sqrt{\fr{(\ell^2+2r_h^2)^2-\ell^4v^2}{1-v^2}}-\ell^2 \; ; \; \; \;
r_h^2 = \frac{1}{2} \left( - \ell^2 + \sqrt{\ell^4 + 4 r_1^2 (r_1^2 + \ell^2)(1-v^2)} \right) \  ,
\ee
and from here it is easy to find the values of $P$ and $\omega$
\ba
P&=&f\gamma v\!\mid_{r=r_1}\;=\; \fr{v r_1^2}{R^2},\\
\omega&=&-\phi\!\mid_{r=r_1}\;=\;
- \sqrt{\frac{(1-v^2) r_1^2}{r_1^2+\ell^2}} \frac{\ell}{R^2} \ .
\ea
Using these results, we then find
\ba
2 \pi \alpha' \pi_{\psi}^0 &=&\fr{ \ell R^2 r_1}{\sqrt{r_1^2(r^2+r_1^2)^2+\ell^4(r^2+v^2r_1^2)+\ell^2((r^2+r_1^2)^2+v^2r_1^4)}},\\
\pi_{x}^0 &=&-\fr{vr^2(r^2(\ell^2+r_1^2+r^2)+\ell^2v^2r_1^2)}{\ell R^2r_1(r^2(\ell^2+r^2)-(1-v^2)r_1^2(\ell^2+r_1^2))}
\sqrt{\fr{\ell^2+r_1^2}{1-v^2}}\pi_{\psi}^0. \label{pix0single}
\ea

To obtain the physical charge density and momentum of the quark on the gauge theory side, we must integrate the above expressions over $r$ from the
horizon, $r_h$, all the way to the flavor brane, which here for simplicity is assumed to lie at $r=\infty$. After some straightforward algebra,
we obtain for the former
\ba
2\pi\alpha' \int_{r_h}^{\infty} dr \, \pi_{\psi}^0&=&i\sqrt{2}\ell R^2r_1\Bigg\{\fr{1}{\sqrt{X_-}}{\rm F}\bigg[{ i}\,{\rm arcsinh}\sqrt{\fr{Z}{X_+}},\fr{X_+}{X_-}\bigg]\nn
&+&{\rm sign}[(X_+-X_-)^2]\bigg(\fr{1}{\sqrt{X_-}}{\rm K}\bigg[\fr{X_+}{X_-}\bigg]-\fr{1}{\sqrt{X_+}}{\rm K}\bigg[\fr{X_-}{X_+}\bigg]\bigg)\Bigg\},
\ea
where K and F are, respectively, the complete and incomplete elliptic integrals of the first kind, and where we have defined
\ba
X_\pm
&\equiv&\ell^4+2\ell^2r_1^2+2r_1^4 \pm \ell\sqrt{\ell^2(\ell^2+2r_1^2)^2-4r_1^2(\ell^2+r_1^2)^2v^2},\\
Z&\equiv&(\ell^2+r_1^2)\(-\ell^2+\sqrt{(\ell^2+2r_1^2)^2-4r_1^2(\ell^2+r_1^2)v^2}\).
\ea
For the latter, the behavior near the horizon on the other hand reveals the logarithmic singularity discussed earlier,
as the expression of $\pi_{x}^0$ in Eq.~(\ref{pix0single})
can be written in the form
\[
2 \pi \alpha' \pi_x^0 = \frac{\mbox{const}}{r-r_h} + {\mathcal O}\left( (r-r_h)^0 \right) \ .
\]

\subsection{Equal charges, no-torque case}

For three equal R charges, $\ell_i=\ell\; \forall i$, it is easy to see that the minimal force condition enforces no constraints on the angles $\theta_i$. We are
therefore free to again choose $\theta_1=\pi/2$, in which case all other angles with the exception of $\psi_1$ decouple. The results of
Eqs.~(\ref{Lag})--(\ref{pix0}) naturally continue to hold, but now we obtain as the relation between the actual and effective horizons
\ba
r_h\(1+\fr{\ell^2}{r_h^2}\)^{3/4}&=&(1-v^2)^{1/4} r_1\(1+\fr{\ell^2}{r_1^2}\)^{3/4}
\ea
and for the expressions of the parameters $P$ and $\omega$
\ba
P&=&f\gamma v\!\mid_{r=r_1}\;=\; \fr{v(\ell^2+r_1^2)}{R^2},\\
\omega&=&-\phi\!\mid_{r=r_1}\;=\; -\fr{\ell \sqrt{1-v^2}}{R^2}\sqrt{1+\fr{\ell^2}{r_1^2}}.
\ea

The conjugate momenta now read
\ba
\pi_{\psi}^0 &=&\ell R^2 r_1r \sqrt{\ell^2+r_1^2}\times\Big(\left(1-v^2\right) \ell^{10}-\left(2 v^2+1\right) r_1^2 \ell^8\nn
&-&r_1^2 \left(\left(4-v^2\right) r^2+\left(1+v^2\right)
   r_1^2\right) \ell^6
   -r^2 r_1^2 \left(r^2-2 \left(1+v^2\right) r_1^2\right) \ell^4\nn
   &+&r^2 r_1^4 \left(4 r^2+\left(v^2+4\right) r_1^2\right) \ell^2+r^2 r_1^4
   \left(r^2+r_1^2\right)^2\Big)^{-1/2},\\
\pi_{x}^0 &=&-\fr{vr_1(\ell^2+r^2)^2}{\ell R^2\sqrt{(1-v^2)(\ell^2+r_1^2)}}\pi_{\psi}^0\nn
&\times& \Big(-\ell^6+3 r^2 r_1^2 \ell^2+v^2 \left(\ell^2+r_1^2\right)^2 \ell^2+r^2 r_1^4+r^4 r_1^2\Big)\\
&\times&
\Big(r^2 v^2 \left(\ell^2+r_1^2\right)^3+\left(r^2-r_1^2\right)
   \left(-\ell^6+r^2 r_1^4+r^2 \left(3 \ell^2+r^2\right) r_1^2\right)\Big)^{-1}.\nonumber
\ea
These expressions are difficult to integrate analytically, but we again observe that only the former is finite (and
integrable) from $r_h$ to infinity, while $\pi_x^0$ has a logarithmic singularity at $r=r_h$.

\subsection{Single charge non-spinning case}

When both $\pi_{\psi_i}^1 \neq 0$ and $\omega_i \neq 0$,
we expect there to be a divergent contribution to the energy
because an infinite amount of work has gone into dragging the string in the angular
$\psi_i$ direction.
The case analyzed in Ref.~\cite{Caceres:2006dj}, where
$\pi_{\psi_i}^1 \neq 0$ and $\omega_i = 0$, will on the other hand not produce a divergent energy contribution
because the torque does no work.  However, it will still produce a divergence in the total
R-charge, as we now show.
In Ref.~\cite{Caceres:2006dj},
two classical solutions for strings rotating in D3-brane backgrounds were found, one corresponding to a polar ($\theta_1=0$) and
the other to an equatorial ($\theta_1=\pi/2$) case. For the first one, one trivially obtains $\pi_{\psi}^0=0$ --- implying that the solution has zero R charge density.
Using (\ref{onechargeP})
and (\ref{onechargeL}), we on the other hand find for the latter
\be
P = \frac{v r_h \sqrt{r_h^2+ \ell^2} }{R^2 \sqrt{1-v^2}} \; ; \; \; \;
L = \ell \sqrt{1-v^2} \ ,
\label{CGPLvalues}
\ee
from which we obtain the non-trivial densities
\ba
{\mathcal L}&=&\sqrt{1-v^2},\\
\pi_{\psi}^0 &=&\fr{ \ell R^2 r_h^2\sqrt{1-v^2}\sqrt{1+\ell^2/r_h^2}}{(r^2+r_h^2+\ell^2)(r^2-r_h^2)},\\
\pi_{x}^0 &=&\fr{vr^2(r^2+\ell^2) }{\sqrt{1-v^2}(r^2+r_h^2+\ell^2)(r^2-r_h^2)}.
\ea
Indeed, there is a logarithmic singularity in the total momentum and R-charge
from the expressions for $\pi_{\psi}^0$ and $\pi_x^0$ at the horizon.

\section{Quasinormal modes and instabilities}
\label{app:rigorous}

The results in this section are logically independent from the rest of the paper
but useful for making the arguments in our stability analysis of the spinning
strings simpler and shorter.  The results are not new\footnote{%
Similar results are reviewed in Ref.~\cite{Landsteiner} for example.
}
 and perhaps not 
so well known.
We wish to show two things.  First, a second order differential operator of the
form
\be
(- \partial_r^2 + P(r) \partial_r + Q(r) ) \psi(r) = \omega^2 F(r)^2 \psi(r) \ ,
\ee
can be rewritten in the form
\be
(-\partial_\rho^2 + V(\rho) ) \phi(\rho) = \omega^2 \phi(\rho) \ .
\label{diffop}
\ee
This result is straightforwardly obtained by redefining
$\psi(r) = e^{h(\rho)} \phi(\rho)$ where $d \rho / dr = F(r)$.
By taking
\[
\partial_\rho h(\rho) = \frac{P F - F'}{2 F^2} \ ,
\]
the first order derivative $\partial_\rho$ in (\ref{diffop}) is forced to vanish
and the potential function becomes
\be
V(\rho) = \frac{1}{F^2} \left( Q + \frac{1}{4} P^2 - \frac{1}{2} P' - \frac{3}{4} \frac{(F')^2}{F^2}
+ \frac{1}{2} \frac{F''}{F} \right) \ ,
\label{potentialfunction}
\ee
where $F' = \partial_r F$ and $P' = \partial_r P$. 

To formulate the next result, we need to make some further assumptions
about the boundary conditions satisfied by $\phi(\rho)$ and the nature
of $V(\rho)$.  We assume that
\[
\lim_{\rho\to -\infty} V(\rho) = 0 \ ,
\]
and that $V(\rho)$ is real, smooth, and finite on the domain $-\infty < \rho < \rho_0$.  
In the $\rho \to -\infty$ limit,
$\phi(\rho)$ becomes a plane wave:  
\[
\lim_{\rho \to -\infty} \phi(\rho) \sim e^{i k x} 
\]
where $k^2  =\omega^2$.  
In this limit, we define a group velocity 
for the plane wave
$v_g  = d\omega / dk = \pm 1$.  
We would like the plane wave to be outgoing, i.e.~to be traveling to more negative $\rho$
and hence choose $v_g = -1$ and $\omega = -k$. 
Given these non-Hermitian boundary conditions, the allowed eigenvalues
of our differential operator will in general be complex.

At the other end of the domain $\rho=\rho_0$, we would like to choose self-adjoint
boundary conditions.  Two simple choices are of course Neumann $\phi'(\rho_0)=0$
and Dirichlet $\phi(\rho_0)=0$.  However in general we need only require that
$\phi'(\rho_0) / \phi(\rho_0)$ be constant.

Given these boundary conditions and assumptions on $V(\rho)$,
we will demonstrate that provided $V(\rho)$ has no bound
states, all the eigenvalues of our second order differential operator
(\ref{diffop}) must lie in the lower complex $\omega$ half plane.
We attempt a proof by contradiction and assume the existence of an
$\omega$ in the upper half plane but no bound states of $V(\rho)$.
Then the outgoing boundary condition implies that the plane wave
$e^{-i\omega x}$ will be exponentially damped at large negative $x$. 
This damping means that the eigenfunction $\phi(\rho)$ is $L^2$
normalizable.  

We know how to analyze this second order differential operator
(\ref{diffop}) for $L^2$ normalizable eigenfunctions.  
With self-adjoint boundary conditions at $\rho=\rho_0$, we have a Schr\"odinger
problem; the operator is clearly self-adjoint and all eigenvalues $\omega^2$
must be real and negative.  Thus the corresponding $\omega$ will lie on the imaginary
axis.  By assumption, this $\omega$ must have positive imaginary part.
But these eigenvalues are precisely the bound states of $V(\rho)$ which we
assumed did not exist.\footnote{%
 We would like to thank L.~Yaffe for helping us understand 
 the analytic properties of these eigenvalues.
 }


\begin{thebibliography}{99}

\bibitem{Herzog:2006se}
  C.~P.~Herzog,
  ``Energy loss of heavy quarks from asymptotically AdS geometries,''
  JHEP {\bf 0609}, 032 (2006)
  [arXiv:hep-th/0605191].


\bibitem{Caceres:2006dj}
  E.~Caceres and A.~Guijosa,
  ``Drag force in charged N = 4 SYM plasma,''
  JHEP {\bf 0611}, 077 (2006)
  [arXiv:hep-th/0605235].


\bibitem{Herzog:2006gh}
  C.~P.~Herzog, A.~Karch, P.~Kovtun, C.~Kozcaz and L.~G.~Yaffe,
  ``Energy loss of a heavy quark moving through N = 4 supersymmetric
  Yang-Mills plasma,''
  JHEP {\bf 0607}, 013 (2006)
  [arXiv:hep-th/0605158].

\bibitem{Gubserdrag}
S.~S.~Gubser,
  ``Drag force in AdS/CFT,''
  Phys.\ Rev.\  D {\bf 74}, 126005 (2006)
  [arXiv:hep-th/0605182].



\bibitem{Gubserspinning}
S.~S.~Gubser,
``Thermodynamics of spinning D3-branes,''
Nucl.\ Phys.\ {\bf B551} (1999) 667-684,
[arXiv:hep-th/9810225].

\bibitem{Behrndt}
K.~Behrndt, M.~Cvetic, and W.~A.~Sabra,
``Non-extreme black holes of five dimensional $N=2$
AdS supergravity,''
Nucl.\ Phys.\ {\bf B553} (1999) 317-332,
[arXiv:hep-th/9810227].

\bibitem{Cai:1998ji}
  R.~G.~Cai and K.~S.~Soh,
  ``Critical behavior in the rotating D-branes,''
  Mod.\ Phys.\ Lett.\  A {\bf 14}, 1895 (1999)
  [arXiv:hep-th/9812121].

\bibitem{Chamblin:1999tk}
  A.~Chamblin, R.~Emparan, C.~V.~Johnson and R.~C.~Myers,
  ``Charged AdS black holes and catastrophic holography,''
  Phys.\ Rev.\  D {\bf 60}, 064018 (1999)
  [arXiv:hep-th/9902170].

\bibitem{CveticGubser}
  M.~Cvetic and S.~S.~Gubser,
  ``Phases of R-charged black holes, spinning branes and strongly coupled
  gauge theories,''
  JHEP {\bf 9904}, 024 (1999)
  [arXiv:hep-th/9902195].

\bibitem{KatzKarch}
A.~Karch and E.~Katz,
  ``Adding flavor to AdS/CFT,''
  JHEP {\bf 0206}, 043 (2002)
  [arXiv:hep-th/0205236].


\bibitem{Cvetic:1999xp}
  M.~Cvetic {\it et al.},
  ``Embedding AdS black holes in ten and eleven dimensions,''
  Nucl.\ Phys.\  B {\bf 558}, 96 (1999)
  [arXiv:hep-th/9903214].


  \bibitem{Clifford}
 T.~Albash, V.~Filev, C.~V.~Johnson, A.~Kundu,
  ``Global Currents, Phase Transitions and
  Chiral Symmetry Breaking in Large $N_c$ Gauge Theory,''
  arXiv:hep-th/0605175.


\bibitem{Gubserwake}
  J.~J.~Friess, S.~S.~Gubser, G.~Michalogiorgakis and S.~S.~Pufu,
  ``The stress tensor of a quark moving through N = 4 thermal plasma,''
  Phys.\ Rev.\  D {\bf 75}, 106003 (2007)
  [arXiv:hep-th/0607022]; \\
  S.~S.~Gubser, S.~S.~Pufu and A.~Yarom,
  ``Sonic booms and diffusion wakes generated by a heavy quark in thermal
  AdS/CFT,''
  arXiv:0706.4307 [hep-th].



\bibitem{Paul}
  P.~M.~Chesler and L.~G.~Yaffe,
  ``The wake of a quark moving through a strongly-coupled plasma,''
  arXiv:0706.0368 [hep-th].


\bibitem{SakaiSugimoto}
T.~Sakai and S.~Sugimoto,
  ``Low energy hadron physics in holographic QCD,''
  Prog.\ Theor.\ Phys.\  {\bf 113}, 843 (2005)
  [arXiv:hep-th/0412141].

\bibitem{Gubserjet}
S.~S.~Gubser,
  ``Jet-quenching and momentum correlators from the gauge-string duality,''
  arXiv:hep-th/0612143.


\bibitem{TeaneySolanajet}
J.~Casalderrey-Solana and D.~Teaney,
  ``Transverse momentum broadening of a fast quark in a N = 4 Yang Mills
  plasma,''
  JHEP {\bf 0704}, 039 (2007)
  [arXiv:hep-th/0701123].

\bibitem{stability}
J.~J.~Friess, S.~S.~Gubser, G.~Michalogiorgakis and S.~S.~Pufu,
  ``Stability of strings binding heavy-quark mesons,''
  JHEP {\bf 0704}, 079 (2007)
  [arXiv:hep-th/0609137]; \\
S.~D.~Avramis, K.~Sfetsos and K.~Siampos,
  ``Stability of strings dual to flux tubes between static quarks in N=4 SYM,''
  Nucl.\ Phys.\  B {\bf 769}, 44 (2007)
  [arXiv:hep-th/0612139]; \\
S.~D.~Avramis, K.~Sfetsos and K.~Siampos,
  ``Stability of string configurations dual to quarkonium states in AdS/CFT,''
  arXiv:0706.2655 [hep-th].

\bibitem{guijosadrag}
M.~Chernicoff and A.~Guijosa,
  ``Energy loss of gluons, baryons and k-quarks in an N = 4 SYM plasma,''
  JHEP {\bf 0702}, 084 (2007)
  [arXiv:hep-th/0611155].

\bibitem{obannontoappear}
A.~O'Bannon,
``Hall Conductivity of Flavor Fields from AdS/CFT,"
arXiv:0708.1994 [hep-th].

\bibitem{Landsteiner}
I.~Amado, C.~Hoyos, K.~Landsteiner and S.~Montero,
  ``Absorption Lengths in the Holographic Plasma,''
  arXiv:0706.2750 [hep-th].

\end{thebibliography}
\end{document}